\documentclass{aa}
\usepackage{graphicx}
\usepackage{natbib}
\usepackage{txfonts}
\begin{document}
%
\def\ltsima{$\; \buildrel < \over \sim\;$}
\def\ltsim{\lower.5ex\hbox{\ltsima}}
\def\gtsima{$\; \buildrel > \over\sim \;$}
\def\gtsim{\lower.5ex\hbox{\gtsima}}
\def\ms{$M_{\odot}$ }
\def\msp{$M_{\odot}$}

   \title{Enrichment history of r-process elements  shaped by a merger of neutron star pairs}
\titlerunning{R-process enrichment by a merger of neutron star pairs}
 \authorrunning{Tsujimoto \& Shigeyama}

   \author{ T. Tsujimoto\inst{1} \and T. Shigeyama\inst{2}}
   

   \institute{\inst{1}National Astronomical Observatory of Japan, Mitaka-shi,
                   Tokyo 181-8588, Japan (\email{taku.tsujimoto@nao.ac.jp}) \\
                   \inst{2}Research Center for the Early Universe, Graduate School of Science, University of Tokyo, 
                   7-3-1 Hongo, Bunkyo-ku, Tokyo 113-0033, Japan}
    

\abstract{
The origin of r-process elements remains unidentified and still puzzles us. The recent discovery of evidence for the ejection of r-process elements from a short-duration $\gamma$-ray burst singled out neutron star mergers (NSMs) as their origin. In contrast, core-collapse supernovae are ruled out as the main origin of heavy r-process elements ($A>$110) by recent numerical simulations. However, the properties characterizing NSM events - their rarity and high yield of r-process elements per event - have been claimed to be incompatible with the observed stellar records on r-process elements in the Galaxy. We add to this picture with our results, which show that the observed constant [r-process/H] ratio in faint dwarf galaxies and one star unusually rich in r-process in the Sculptor galaxy agree well with this rarity of NSM events. Furthermore, we found that a large scatter in the abundance ratios of r-process elements to iron in the Galactic halo can be reproduced by a scheme that incorporates an assembly of various protogalactic fragments, in each of which r-process elements supplied by NSMs pervade the whole fragment while supernovae distribute heavy elements only inside the regions swept up by the blast waves. Our results demonstrate that NSMs occurring at Galactic rate of 12-23 Myr$^{-1}$ are the main site of r-process elements, and we predict the detection of gravitational waves from NSMs at a high rate with upcoming advanced detectors.
}


 \maketitle
%
 
\section{Introduction}

Where is the origin of r-process elements, and how has the enrichment proceeded through cosmic time? Clear answers to these questions continue to elude us. Since r-process nucleosynthesis demands an extremely neutron-rich environment, the possible astrophysical sites are limited to two events: core-collapse supernovae (CCSNe) as deaths of massive stars that leave behind neutron stars  (NSs) or NS mergers (NSMs) \citep[e.g.,][]{Truran_81}. Until very recently, the only information available to identify the origin of r-process nucleosynthesis from an observational aspect were records of r-process elements imprinted on spectral lines of long-lived stars in the Galaxy. The enrichment history of r-process elements in these old stars seemed to support CCSNe as their source and to preclude the NSM scenario. The two features predicted by the NSM scenario - a delayed enrichment and a large scatter in r-process/Fe ratios - have thus far been claimed to be incompatible with observed features \citep{Mathews_90, Cowan_91, Argast_04}. A delayed enrichment has been suggested to be an unavoidable result of the rarity of NSM events and the long time-scale of the orbital decay of an NS binary because of gravitational radiation. In addition, the high yield of r-process elements per NSM probably results in stars that are extremely rich in r-process products, far beyond the observed level \citep{Argast_04}. Accordingly, most theoretical work related to the r-process origin has focused on determining the production of r-process elements from CCSNe \citep[e.g.,][]{Woosley_94, Winteler_12}. 

However, the tide has changed from both theoretical and observational perspectives. Recent numerical simulations of CCSNe revealed that the wind blown from a nascent NS becomes not as neutron-rich as assumed because of interactions with neutrinos and is therefore unsuitable for r-process nucleosynthesis \citep{Thompson_01, Wanajo_13}. In addition, the recent detection of a near-infrared light in the afterglow of a short-duration $\gamma$-ray burst GRB 130603B at day 9 confirmed the prediction of the theoretical model of a kilonova \citep{Barnes_13, Tanvir_13}. A kilonova is an event bright in near-infrared light emitted by matter with a mass of $\sim 0.01$ \ms composed of almost pure r-process elements that are ejected more or less isotropically as a result of merging NSs \citep{Barnes_13, Hotokezaka_13}, while the associated GRB is thought to be a result of collimated jets moving toward us. The expansion velocities are expected to be 10\% to 30\% of the speed of light ($0.1c$-$0.3c$), depending on the adopted equation of state for dense matter. This striking discovery together with failures to produce r-process elements in SNe naturally drives our growing attention to the possibility that the NSM might be a major site of r-process elements. To investigate this in depth, we studied the r-process enrichment in galaxies to provide new insights into both observation and theory. We present our results in the following two  sections.

\section{r-process enrichment in dwarf galaxies}
   \begin{figure}[htb]
   \centering
  \includegraphics[angle=0,width=0.35\textwidth]{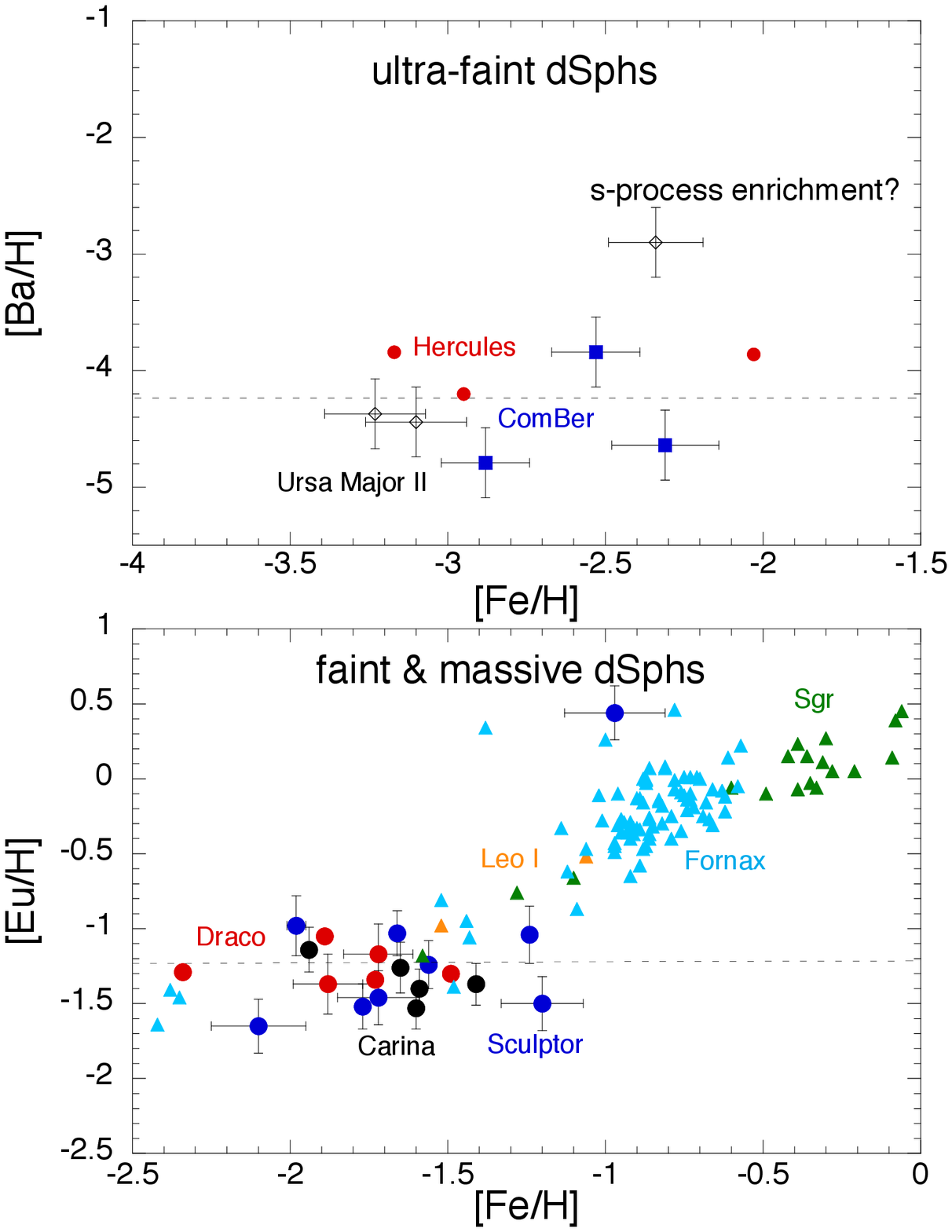}
       \caption{Observed correlations of Eu/H (Ba/H) with Fe/H for ultra-faint (least massive) dwarf dSph galaxies (upper panel) and faint (less massive: circles) and massive (triangles) dSph galaxies (lower panel). The observed data are assembled from the literature \citep{Frebel_10, Koch_13, Shetrone_01, Shetrone_03, Geisler_05, Cohen_09, Letarte_10, McWilliam_13}. We see no broad increase in the Eu (or Ba: no measurements of the element Eu for ultra-faint dSphs) abundance for less massive dSphs. The average [Eu(Ba)/H] are shown by dashed lines. The high Ba abundance for one star of Ursa Major II can be ascribed to the enrichment by s-process.  In the lower panel, we exclude the data of Ursa Minor, which has a compatible feature  \citep{Shetrone_01}, but is not supported by \citet{Cohen_10}.
              }
         \label{}
   \end{figure}

Beyond the Galaxy, the detailed abundance data accumulated to date from nearby dwarf galaxies have a high potential to provide a clue on the origin of r-process elements. The Galaxy is surrounded by tens of dwarf galaxies spanning over six orders of magnitude in stellar mass, from ultra-faint dwarf spheroidal (dSph) galaxies to massive dwarf galaxies. The NSM origin scenario predicts that the r-process enrichment takes a clearly different track in galaxies with small mass scales, which is a result from the low rate of NSMs, $<4\times10^{-15}$ events per year per solar mass \citep{Lorimer_08} inferred from observations for binary pulsars. For example, an ultra-faint dSph galaxy with a stellar mass of about $10^4$ \ms is expected to have undergone $\ll$ 0.1 events in its past, implying no enrichment of r-process elements. Figure 1 shows that the correlations of ratios of a representative r-process element Eu (Ba) to H against Fe/H ratios for local dSph galaxies can be classified into three different trends according to the mass scales. We see no clear signature of r-process enrichment for less massive dSph galaxies. Their broadly constant Eu (Ba) abundances are considered to be a reflection of a pre-enriched gas that has originally resided in a more massive reservoir. One exceptional star that exhibits unusually high [Eu/H] at [Fe/H]$\sim-1$ in the Sculptor dSph might be a result of an emergence of NSM at its late stage of chemical evolution \citep[see also][]{2009ARA&A..47..371T}.

On the other hand, for dSph galaxies more massive than $\sim10^7$\msp, the Eu abundance is clearly shown to increase in accordance with the Fe abundance. The Large Magellanic Cloud obeys the same correlation \citep{Swaelmen_13}. Since its slope is determined by the ratio of the production rates between Fe and Eu, we can deduce an NSM rate as one per $\sim1000$ core-collpase SNe, assuming the typical Fe yield from a CCSN of 0.1 \ms \citep{Hamuy_03} and the typical r-process element yield of 0.01 \ms from an NSM \citep{Barnes_13, Hotokezaka_13}. Then, this NSM rate combined with the total luminosities plus a canonical stellar initial mass function (IMF) allocates the total number of NSM events of 0.03 to Coma Berenices dSph (5,000$L_\odot$) and 2 to Carina dSph ($4\times10^5 L_\odot$), for instance. Considering the large uncertainty in this estimate including an ambiguity of the IMF that affects the stellar mass by a factor of $\sim$ 2 \citep{Martin_08}, it is no wonder that no NSM has occurred in these less massive dSph galaxies. From a different point of view, the observed feature of constant [Eu/H] in faint dSphs supports the ejected mass of NSM of $\sim$0.01 \msp, and disfavors masses as small as $1\times10^{-3}$ \msp, which corresponds to the lower limit predicted by numerical simulations \citep{Bauswein_13} and would result in a ten times higher rate of NSM events, that is, one per 100 CCSNe. 

\section{Propagation of NSM ejecta}

A recent model developed to account for new results from observations of short-duration GRBs has dramatically shortened the merging time to a time-scale ranging from 0.001 to 0.1 Myr \citep{Belczynski_06}, although it was first suggested that NSs in a binary system merge in 100-1000 Myr after birth \citep{Portegies_98}. When an NSM occurs in a protogalactic fragment after the less massive progenitor explodes as a supernova, the ejected r-process nuclei that recede from the merging site reach asymptotic speeds of $>0.1c$ in $\sim 10$ ms \citep{Bauswein_13}. These nuclei probably do not dissipate the kinetic energies in the interstellar matter (ISM) because the stopping length $l_{\rm s}$ of a $^{153}$Eu nucleus (as a representative r-process element) at the speed of 0.2$c$ is estimated  to be $l_{\rm s}\sim400/n$ kpc from the ionization-loss rate given in \citet{2002cra..book.....S} ($n$ denotes the mean number density of neutral hydrogen in the ISM in cgs units), much longer than the size of a proto-galactic fragment. Here the ionization-loss rate  $w(\beta)$ (MeV/s) for a nucleus with the mass number $A$ and charge $Z$ is expressed as a function of the velocity $v=c\beta$, 
$w(\beta)={1.82\times10^{-13} (0.0185 \beta+1) n Z^2\left\{1-1.034\exp\left(-137\beta Z^{-0.688}\right)\right\}^2}/{ A \beta}.$
According to \citet{Tegmark_97}, a fragment with a mass of  $10^5$ \ms has an average number density of $\sim 20$ cm$^{-3}$, more massive fragments have higher number densities. Therefore the interaction of the ejected r-process nuclei with the ISM cannot be treated in the same way as the fluid approximation.

Alternatively, turbulent magnetic fields on the order of $\mu$G, which is a typical value observed for the Galactic ISM, are sufficient to help such a nucleus lose a significant fraction of the energy through ionization inside the fragment as long as the typical coherence length $l_{\rm c}$ of the field does not exceed $R_{\rm p}^2/l_{\rm s}\sim0.03$ pc $n$($R_{\rm p}/100$ pc)$^2$, where $R_{\rm p}$ denotes the scale of the fragment. The typical value of $l_{\rm c}$ is about 100 AU \citep{Houde_11}.  Thus r-process elements ejected from a NSM diffuse over the entire fragment on a timescale of $l_{\rm s}/0.2c\sim7/n$ Myr. As a consequence, an NSM can enrich a protogalactic fragment with r-process elements in less than 10 Myr after the birth of the progenitor stars. Such physical processes are far different from those for heavy elements that are ejected from a CCSN that expands at a speed of a few thousand km/s. Since the stopping length due to ionization is roughly proportional to the fourth power of the velocity,  a $^{56}$Fe ($^{16}$O) nucleus moving at this order of speed is found to lose its energy after traveling a distance of only $< 100/n$ pc ($< 20/n$ pc) if the same formula from \citet{2002cra..book.....S} is applied. Thus heavy elements ejected from an SN are expected to be trapped inside the ISM swept up by the blast wave. Note that the propagation of NSM ejecta was previously treated in the same way as for the ejecta of SNe \citep{Argast_04}. 

If the speeds of r-process elements exceed $0.2c$, the nuclei have energies higher than the threshold for spallation reactions with protons in the ISM \citep[see][]{Carlson_96}. As a result of the large cross sections, the r-process elements suffer from spallation reactions before losing the energies through ionization. The abundance pattern of r-process elements in the ejecta of NSM calculated by  \citet{Bauswein_13} exhibits a too prominent third peak, or too low first two peaks compared with the observed abundance pattern \citep{Burris_00}. This may be reconciled by including yields in a neutrino-driven wind where low-A elements are synthesized \citep{Rosswog_14}. Whether spallation reactions are needed to reproduce the observed abundance pattern of r-process elements might give a clue to distinguish the proposed equations of state for dense matter because the equation of state is a crucial factor in determining the ejecta speeds \citep{Bauswein_13}. 

\section{Chemical evolution of r-process elements}

\subsection{Galactic halo}
Our proposed interplay between the ejecta of NSM and the ISM gives a crucial key for understanding the feature of r-process elements in the Galactic halo stars that are characterized by a large scatter over three orders of magnitude in their ratios with respect to Fe. The observed large scatter can be ascribed to various mass scales of the protogalactic fragments that give birth to halo field stars \citep{Searle_78}. For example, one NSM event enriches the entire gas fragment with r-process elements to [Eu/H]=$-0.5$ for a mass of $10^6$ \msp, but only up to [Eu/H]=$-3.5$ for the $10^9$ \ms fragment. Although the probability of having NSMs in each small fragment is low, many small fragments are expected to form in a halo progenitor. Some of these small fragments probably  host NS binaries that have coalesced immediately after the first SN events, and thus we naturally expect low-metallicity stars that exhibit [Eu/Fe]$\sim+2$, as observed. Furthermore, an IMF favoring massive stars for [Fe/H]$<-2$ proposed from the statistics of carbon stars \citep{Suda_13} significantly increases the probability of an early emergence of NSM. 

Based on the above scenario, we calculated the chemical evolution of the halo, incorporating an SN-induced star formation into the models, which necessarily causes a chemical inhomogeneity among halo stars, as observed \citep{Tsujimoto_99}. The chemical characteristics of halo stars do not support the idea that the stars were formed from well-mixed gas. Progress in inventing the theoretical scheme of star formation has been made by close attention to the abundance patterns of heavy elements of metal-poor halo stars \citep{McWilliam_95, Cayrel_04} that possess fossil records of star formation in an early Galaxy. From analyses of abundance patterns of metal-poor halo stars with [Fe/H]$<-2.5$, it has ben found that heavy elements ejected from SNe were only mixed into the gas swept up by individual SNe in the halo \citep{Shigeyama_98}. In our model, stars are assumed to have formed from dense shells swept up by individual SNe and thus inevitably inherited their elemental abundance patterns from the dense shells. Massive stars among them eventually exploded as SNe, which again triggered the formation of new stars. This SN-induced star formation proceeded for generations until SN remnants became unable to sweep up a sufficient amount of gas to form dense shells. The mass fraction of the shell of each SN explosion turning into stars, which controls the star formation efficiency, is assumed to be a constant value of 0.004 \citep{Tsujimoto_99}.

This scheme for star formation is combined with two critical hypotheses to trace the chemical evolution of r-process elements in the halo. First, we assumed that the production site of r-process elements are NSM events, each of which ejects r-process-rich matter of the mass of 0.01 \ms with a time delay of 10-30 Myr \citep{Belczynski_06}. For each fragment, we furthermore assumed that the ejected r-process element is mixed with the entire ISM and the event rate of NSM, to which a frequency 1000 times lower than that of CCSNe is basically assigned, is proportional to the fragment mass. Second, we adopted the view that has first been  proposed by \citet{Searle_78}: that the halo was formed through accretion of protogalactic fragments associated with small dark matter haloes as Galactic building blocks. Note that this picture is almost established by the current scheme of hierarchical galaxy formation scenario and has previously been applied to discuss the chemical properties of the Galactic halo \citep{Travaglio_01, Komiya_14}.
   \begin{figure}[htb]
   \centering
  \includegraphics[angle=0,width=0.35\textwidth]{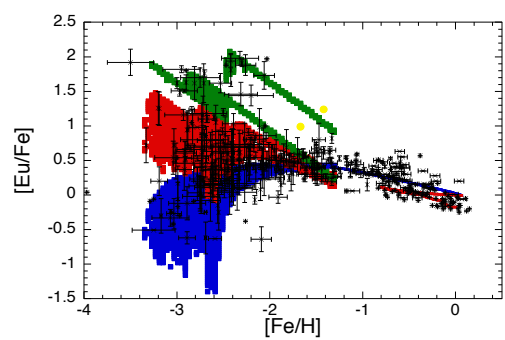}
       \caption{Enrichment history of the r-process element Eu in the Galaxy predicted by our models of chemical evolution, into which NSM events are incorporated as the production site of r-process elements. The results are compared with the observed Galactic data in the solar vicinity denoted by crosses \citep{Suda_08}. Two stars marked by yellow circles (belonging to the Galactic bulge/the Ursa Minor dSph) are added to the figure; they exhibit unusually high Eu/Fe ratios at moderate metallicities \citep{Johnson_13}. The Eu/Fe-Fe/H correlations for the Galactic halo that result from our model are represented by three typical components with different colors. The three components correspond to stars born from protogalactic fragments with different masses - small (green), moderate (red), and massive (blue). In addition, the calculated evolutionary paths for the Galactic thick and thin disks are shown by the blue and red curves, respectively.
              }
         \label{}
   \end{figure}

The results are shown in Figure 2 with three different color-coded clouds in the Eu/Fe-Fe/H diagram, each of which corresponds to a different mass-scale of protogalactic fragments. Two separate green components consist of stars from two fragments with gaseous masses of $2\times10^6$ \ms (lower part) and $2\times10^5$ \ms (upper part). It is assumed that the former hosts two NSM events with one at a very early epoch and the latter only one with a time delay of about two hundred Myr. Since a $2\times10^5$ \ms fragment is expected to host 0.1 event on average, an early emergence of NSM in similar-sized fragments is very unlikely, resulting in the absence of halo stars exhibiting [Eu/Fe]$>+2.5$. Some stars show unusually high Eu/Fe ratios at moderate metallicities, the origin of which remains a mystery \citep[yellow circles;][]{Johnson_13}. Because the Eu/Fe ratio at each Fe/H is inversely proportional to the fragment mass, these stars have probably been born in the smaller fragments and subsequently merged with the more massive present-day systems that host relatively low Eu/Fe stars. On the other hand, the blue component corresponds to a massive fragment of $\sim10^9$ \ms where NSM events steadily occur at a rate of one per 1000 CCSNe. The red component is middle between the two -a superposition of stars from two fragments with masses of $10^7$ \ms (with five NSM events) and $2\times10^7$ \ms (with ten events). In addition, some halo stars apparently originate from very small fragments of $<10^5$\ms that do not host NSMs. These stars are likely to become mixed in the red or blue cloud depending on the level of pre-enrichment inside their  fragments, as implied by local faint dSphs. In individual fragments, about 10 \% of the gas is converted into stars for a period of about 500 Myr to match the observed metallicity distribution of halo stars \citep{Tsujimoto_99}. 

We assume that the mass distribution of fragments obeys the mass  function of substructures in the simulated Galaxy and checked whether the statistics of the fragments are consistent with the observed populations of stars in the green and red clouds of Figure 2. The $10^6$\ms and $10^9$\ms gaseous fragments correspond to the progenitors of a less massive dSph (lower green) and an SMC-sized galaxy (red), respectively. Figure 2 shows the observed number ratio of halo star population in individual colored (green, blue) clouds of $\sim$1:6. On the other hand, simulations for the Galaxy formation predict the corresponding ratio to be $\sim$1:10 \citep{Moore_99}. The estimates from two independent methods agree well. We stress that a large scatter in the  abundance ratio unique to the r-process elements in the Galactic halo is an early relic of a structure formation via mass assembly of small building blocks predicted by an established modern paradigm in the cold dark matter Universe. 

\subsection{Galactic disks}

A remarkable decrease in scatter of the Eu/Fe ratios with increasing Fe abundance enables us to apply one-zone models to the chemical evolution for the Fe-rich ([Fe/H]$>-1.5$) Galactic components, that is, the disk. The disk was modeled separately for the thick and thin disk scenario, and each modeling was the same as in our previous works \citep{Tsujimoto_12}. In this scheme, the thick disk was first formed rapidly, enriched with Fe from [Fe/H] = $-1.5$ up to [Fe/H]$\sim+0$ by CCSNe as well as Type Ia SNe. Then the thin disk was gradually formed from the thick disk's debris gas that accreted metal-poor gas from the halo. The metal in the debris gas was diluted by the accretion. As a result, the metallicity [Fe/H] decreased to $-0.8$. This reverse evolution came to an end when the metal supply from star formation overcame the dilution effect. Then the thin disk subsequently resumed the metal enrichment. Note that the feature of both thick and thin disks might also be explained by the radial migration \citep{Schonrich_09}. We here adopted an NSM rate of one per 2000 CCSNe to obtain a best fit to the observed data, together with the same yield and delay time for each merger event as in modeling the halo. 

\section{Prospects}

The present study will be continued with the following future works: first, we need to accumulate more data of the Eu abundance across a wide range of [Fe/H] for individual faint dSph galaxies. In particular, we need to confirm whether the Sculptor dSph exhibits a sudden increase in [Eu/H] at [Fe/H]$\sim -1$. Second, the fission of nuclei with $A>230$ needs to be explored to fully understand the abundance pattern immediately after the production in NSMs. Furthermore, it will be an intriguing task to obtain the pattern of r-process elements modified by spallation through the propagation in the ISM after the production in NSMs and to compare them with the observed patterns for metal-poor halo stars \citep[e.g.,][]{Roederer_10}.  

Our analysis that identified NSM events as the dominant production site for r-process elements yields an NSM rate of one per $\sim$ 1000-2000 CCSNe, which is equivalent to the Galactic merger rate of $\sim$12-23 Myr$^{-1}$ using the present-day Galactic CCSN rate of 2.3 SNe per century \citep{Li_11}. It matches the range $8^{+9}_{-5}$ Myr$^{-1}$ of the rate deduced from currently known double neutron star binaries \citep{Kim_03} \citep[but see other estimates such as $118^{+174}_{-79}$ Myr$^{-1}$ by][or 1-1000 Myr$^{-1}$ by Abadie et al.~ 2010]{Lorimer_08}. This coincidence of the results from two independent methods increases the possibility that an NSM is the main site of r-process elements and promises a coming era of gravitational wave astronomy at a high detection rate by KAGRA, VIRGO, and LIGO. The advanced LIGO-VIRGO network, for instance, will catch a first signal in 2018-2019 and finally detect $\sim$ 5-10 NSM events annually \citep[cf.][]{Abadie_10}.



\begin{thebibliography}{}

\bibitem[Abadie et al.(2010)]{Abadie_10}
Abadie, J., Abbott, B. P., Abbott, R., et al. 2010, CQGra, 27, 173001
\bibitem[Argast et al.(2004)]{Argast_04}
Argast, D., Samland, M., Thielemann, F.-K., \& Qian, Y.-Z. 2004, A\&A, 416, 997
\bibitem[Barnes \& Kasen(2013)]{Barnes_13}
Barnes, J., \& Kasen, D. 2013, \apj, 775, 18
\bibitem[Bauswein et al.(2013)]{Bauswein_13}
Bauswein, A., Goriely, S., \& Janka, H.-T. 2013, \apj, 773, 78
\bibitem[Belczynski et al.(2006)]{Belczynski_06}
Belczynski, K., Perna, R., Bulik, T., Kalogera, V., Ivanova, N., \& Lamb, D. Q. 2006, \apj, 648, 1110
\bibitem[Burris et al.(2000)]{Burris_00}
Burris, D. L., Pilachowski, C. A., Armandroff, T. E., et al. 2000, \apj, 544, 302
\bibitem[Carlson(1996)]{Carlson_96}
Carlson, R. F. 1996, Atomic Data and Nuclear Data Tables, 63, 93
\bibitem[Cayrel et al.(2004)]{Cayrel_04}
Cayrel, R., Depagne, E., Spite, M., et al. 2004, A\&A, 416, 1117
\bibitem[Cohen \& Huang(2010)]{Cohen_10}
Cohen, J. G., \& Huang, W. 2010, \apj, 719, 931
\bibitem[Cohen \& Huang(2009)]{Cohen_09}
Cohen, J. G., \& Huang, W. 2009, \apj, 701, 1053
\bibitem[Cowan et al.(1991)]{Cowan_91}
Cowan, J. J., Thielemann, F.-K., \& Truran, J. W. 1991, Physics Reports, 208, 267
\bibitem[Frebel et al.(2010)]{Frebel_10}
Frebel, A., Simon, J. D., Geha, M., \& Willman, B. 2010, \apj, 708, 560
\bibitem[Geisler et al.(2005)]{Geisler_05}
Geisler, D., Smith V. V., Wallerstein G., Gonzalez, G., \& Charbonnel C. 2005, AJ, 129, 1428  
\bibitem[Hamuy(2003)]{Hamuy_03}
Hamuy, M. 2003, \apj, 582, 905
\bibitem[Hotokezaka et al.(2013)]{Hotokezaka_13}
Hotokezaka, K., Kyutoku, K., Tanaka, M., et al. 2013, \apj, 778, L16
\bibitem[Houde et al.(2011)]{Houde_11}
Houde, M.,  Rao, R., Vaillancourt, J. E., \& Hildebrand, R. H. 2011, \apj, 733, 109
\bibitem[Johnson et al.(2013)]{Johnson_13}
Johnson, C. I., McWilliam, A., \& Rich, R. M. 2013, \apj, 775, L27
\bibitem[Kim et al.(2003)]{Kim_03}
Kim, C., Kalogera, V., \& Lorimer, D. R. 2003, \apj, 584, 985
\bibitem[Koch et al.(2013)]{Koch_13}
Koch, A., Feltzing, S., Aden, D., \& Matteucci, F. 2013, A\&A, 554, A5
\bibitem[Komiya et al.(2014)]{Komiya_14}
Komiya, Y., Yamada, S., Suda, T., \& Fujimoto, M. Y. 2014, \apj, 783, 132
\bibitem[Letarte et al.(2010)]{Letarte_10}
Letarte, B., Hill, V., Tolstoy, E., et al. 2010,  A\&A, 523, A17
\bibitem[Li et al.(2011)]{Li_11}
Li, W., Chornock, R., Leaman, J., et al. 2011, MNRAS, 412, 1473
\bibitem[Lorimer(2008)]{Lorimer_08}
Lorimer, D. R. 2008, LRR, 11, 8
\bibitem[Martin et al.(2008)]{Martin_08}
Martin, N. F., de Jong, J. T. A., \& Rix, H.-W. 2008, \apj, 684, 1075
\bibitem[Mathews \& Cowan(1990)]{Mathews_90}
Mathews, G. J., \& Cowan, J. J. 1990, Nature, 345, 491
\bibitem[McWilliam et al.(2013)]{McWilliam_13}
McWilliam, A., Wallerstein, G., \& Mottini, M. 2013, \apj, 778, 149
\bibitem[McWilliam et al.(1995)]{McWilliam_95}
McWilliam, A., Preston, G. W., \& Sneden, C. 1995, AJ, 109, 2757
\bibitem[Moore et al.(1999)]{Moore_99}
Moore, B., Ghigna, S., Governato, F., Lake, G., Quinn, T., Stadel, J., \& Tozzi, P. 1999, ApJ, 524, L19
\bibitem[Portegies Zwart \& Yungelson(1998)]{Portegies_98}
Portegies Zwart, S. F., \& Yungelson, L. R. 1998, A\&A, 332, 173
\bibitem[Roederer et al.(2010)]{Roederer_10}
Roederer, I. U., Cowan, J. J., Karakas, A. I., et al. 2010, \apj, 724, 975
\bibitem[Rosswog et al.(2014)]{Rosswog_14} 
Rosswog, S., Korobkin, O., Arcones, A., Thielemann, F.-K., \& Piran, T.\ 2014, \mnras, 439, 744 
\bibitem[Searle \& Zinn(1978)]{Searle_78}
Searle, L., \& Zinn, R. 1978, \apj, 225, 357
\bibitem[Schlickeiser(2002)]{2002cra..book.....S} 
Schlickeiser, R.\ 2002, Cosmic ray astrophysics (Berlin: Springer)
\bibitem[Sch\"{o}nrich \& Binney(2009)]{Schonrich_09}
Sch\"{o}nrich, R., \& Binney, J. 2009, MNRAS, 399, 1145
\bibitem[Shetrone et al.(2003)]{Shetrone_03}
Shetrone, M., Venn, K. A., Tolstoy, E., Primas, F., Hill, V., \& Kaufer, A. 2003, AJ, 125, 684
\bibitem[Shetrone et al.(2001)]{Shetrone_01}
Shetrone, M., Cote, P., \& Sargent, W. L. W. 2001, \apj, 548, 592
\bibitem[Shigeyama \& Tsujimoto(1998)]{Shigeyama_98}
Shigeyama, T., \& Tsujimoto, T. 1998, \apj, 507, L135
\bibitem[Suda et al.(2013)]{Suda_13}
Suda, T., Komiya, Y., Yamada, S., et al. 2013, MNRAS, 432, L46
\bibitem[Suda et al.(2008)]{Suda_08}
Suda, T., Katsuya, Y., Yamada, S., et al. 2008, PASJ, 60, 1159
\bibitem[Tanvir et al.(2013)]{Tanvir_13}
Tanvir, N. R., Levan, A. J., Fruchter, A. S., et al. 2013, Nature, 500, 547
\bibitem[Tegmark et al.(1997)]{Tegmark_97}
Tegmark, M., Silk, J., Rees, M. J., Blanchard, A., Abel, T., \& Palla, F. 1997, \apj, 474, 1
\bibitem[Thompson et al.(2001)]{Thompson_01}
Thompson, T. A., Burrows, A., \& Meyer, B. S. 2001, \apj, 562, 887
\bibitem[Tolstoy et al.(2009)]{2009ARA&A..47..371T} Tolstoy, E., Hill, V., \& Tosi, M.\ 2009, \araa, 47, 371 
\bibitem[Travaglio et al.(2001)]{Travaglio_01}
Travaglio, C., Galli, D., \& Burkert, A. 2001, \apj, 547, 217
\bibitem[Tsujimoto \& Bekki(2012)]{Tsujimoto_12}
Tsujimoto, T., \& Bekki, K. 2012, \apj, 747, 125
\bibitem[Tsujimoto et al.(1999)]{Tsujimoto_99}
Tsujimoto, T., Shigeyama, T., \& Yoshii, Y. 1999, \apj, 519, L63
\bibitem[Truran(1981)]{Truran_81}
Truran, J. W. 1981, A\&A, 97, 391
\bibitem[Van der Swaelmen et al.(2013)]{Swaelmen_13}
Van der Swaelmen, M., Hill, V., Primas, F., \& Cole, A. A. 2013,  A\& A, 560, A44
\bibitem[Wanajo(2013)]{Wanajo_13}
Wanajo, S. 2013, \apj, 770, L22
\bibitem[Winteler et al.(2012)]{Winteler_12} 
Winteler, C., K{\"a}ppeli, R., Perego, A., et al.  2012, \apjl, 750, L22 
\bibitem[Woosley et al.(1994)]{Woosley_94} 
Woosley, S. E., Wilson, J. R., Mathews, G. J., Hoffman, R. D., \& Meyer, B. S. 1994, \apj, 433, 229 
\end{thebibliography}
\end{document}